\documentclass[%
 reprint,
 amsmath,amssymb,
 aps,
]{revtex4-2}

\usepackage{graphicx}
\usepackage{dcolumn}
\usepackage{bm}
\usepackage{hhline}
\usepackage{array}

\begin{document}



\title{Surface Acoustic Wave Cavity Optomechanics with WSe$_2$ Single Photon Emitters}

\author{Sahil D. Patel}
\altaffiliation{These authors contributed equally to this work.}
\affiliation{Department of Electrical and Computer Engineering, University of California Santa Barbara, Santa Barbara, CA 93106, USA}
\author{Kamyar Parto}
\altaffiliation{These authors contributed equally to this work.}
\affiliation{Department of Electrical and Computer Engineering, University of California Santa Barbara, Santa Barbara, CA 93106, USA}
\author{Michael Choquer}
\altaffiliation{These authors contributed equally to this work.}
\affiliation{Department of Electrical and Computer Engineering, University of California Santa Barbara, Santa Barbara, CA 93106, USA}
\author{Sammy Umezawa, Landon Hellman, Daniella Polishchuk}
\affiliation{Department of Electrical and Computer Engineering, University of California Santa Barbara, Santa Barbara, CA 93106, USA}

\author{Galan Moody}
\email{moody@ucsb.edu}
\affiliation{Department of Electrical and Computer Engineering, University of California Santa Barbara, Santa Barbara, CA 93106, USA}

\date{\today}

\begin{abstract}
\noindent Surface acoustic waves (SAWs) are a versatile tool for coherently interfacing with a variety of solid-state quantum systems spanning microwave to optical frequencies, including superconducting qubits, spins, and quantum emitters. Here, we demonstrate SAW cavity optomechanics with quantum emitters in 2D materials, specifically monolayer WSe$_2$, on a planar lithium niobate SAW resonator driven by superconducting electronics. Using steady-state photoluminescence spectroscopy and time-resolved single-photon counting, we map the temporal dynamics of modulated 2D emitters under coupling to different SAW cavity modes, showing energy-level splitting consistent with deformation potential coupling of 30 meV/$\%$. We leverage the large anisotropic strain from the SAW to modulate the excitonic fine-structure splitting on a nanosecond timescale, which may find applications for on-demand entangled photon-pair generation from 2D materials. Cavity optomechanics with SAWs and 2D quantum emitters provides opportunities for compact sensors and quantum electro-optomechanics in a multi-functional integrated platform that combines phononic, optical, and superconducting electronic quantum systems.
\end{abstract}

\maketitle


\section{Introduction}
The coupling between solid-state quantum emitters and confined acoustic modes in optomechanical resonators is an elegant approach for coherently controlling, transferring, and entangling a variety of quantum degrees of freedom, including photons, phonons, and spins \cite{Delsing2019,Schuetz2015,choquer2022quantum}. Among the various optomechanical systems under development \cite{Safavi-Naeini2019}, surface acoustic wave (SAW) resonators integrated with single-photon emitters (SPEs) have several potential advantages over other strategies. SPEs are remarkably sensitive to local strain and exhibit frequency shifts nearly two orders-of-magnitude larger ($\sim$10 GHz/pm) than microscale optical resonators ($\sim$100 MHz/pm). The use of SPEs provides a strong optical nonlinearity that ensures only single photons are emitted, typically with sub-nanowatt optical power requirements. Heterogeneous integration schemes have been used to combine several types of SPEs with SAW resonators, including III-V quantum dots (QDs) \cite{Gell2008,Metcalfe2010,nysten2020hybrid,decrescent2022large} and neutral divacancy centers in SiC \cite{Whiteley2019}. When operated in the sideband-resolved regime using gigahertz-frequency resonators, parametric modulation of SPEs enables microwave-frequency information to be encoded as optical photons, which may pave the way for efficient and low-noise transduction between microwave and optical frequency qubits.

Outstanding challenges with existing strategies for SPE optomechanics include the complexity in materials growth and device fabrication, weak optomechanical coupling relative to the emitter and SAW resonator decoherence rates, and the use of weakly piezoelectric materials such as GaAs and ZnO. The recent discovery of SPEs in 2D materials \cite{azzam2021prospects,kianinia2022quantum}, such as WSe$_2$ \cite{Srivastava2015,he2015single,chakraborty2015voltage,Koperski2015,tonndorf2015single} and hexagonal boron nitride \cite{Tran2016,jungwirth2016temperature}, provide an opportunity to address each of these challenges. 2D SPEs, which originate from crystalline defects in the host material, exhibit high optical extraction efficiency and brightness with detection rates up to 25 MHz \cite{grosso2017tunable,luo2018deterministic,zhao2021site}, indistinguishable \cite{fournier2022two} and near transform-limited linewidths \cite{dietrich2020solid}, high single-photon purity, unique spin-valley phenomena \cite{schaibley2016valleytronics,exarhos2019magnetic,stern2022room}, high working temperatures \cite{tran2016robust,luo2019single,parto2021defect}, and site selective engineering \cite{branny2017deterministic,palacios2017large,fournier2020position,parto2022cavity}. The layered structure of 2D materials arising from van der Waals forces ensures that the defects are two-dimensional and are able to function at surfaces, devoid from any surface states, allowing for strong proximity interaction with their surrounding environment. 

This strong proximity interaction, in addition to the site specific fabrication and relaxed lattice-matching requirements, make 2D SPEs an ideal two-level system to be integrated with SAW resonators. Proximity effects allow for efficient coupling with the SAW deformation potential, while the ability to deterministically transfer 2D monolayers onto nearly any surface allows for nanoscale precision in positioning of a single SPE within the resonator. Indeed, modulation of few-layer hBN SPEs with propagating SAWs has resulted in strong deformation potential coupling \cite{Lazi2019}. Amongst 2D material SPEs, WSe$_2$ emitters are more suitable for cavity optomechanics due to larger proximity effects in monolayers, which partially contributes to their large strain susceptibility up to 120 meV/$\%$ \cite{iff2019strain}. The integration of 2D SPEs with high-quality SAW resonators, which has not been demonstrated to date, is the next critical step for exploring the potential of 2D materials for quantum optomechanics.  

\begin{figure*}[t!] \centering
     \includegraphics[scale=1]{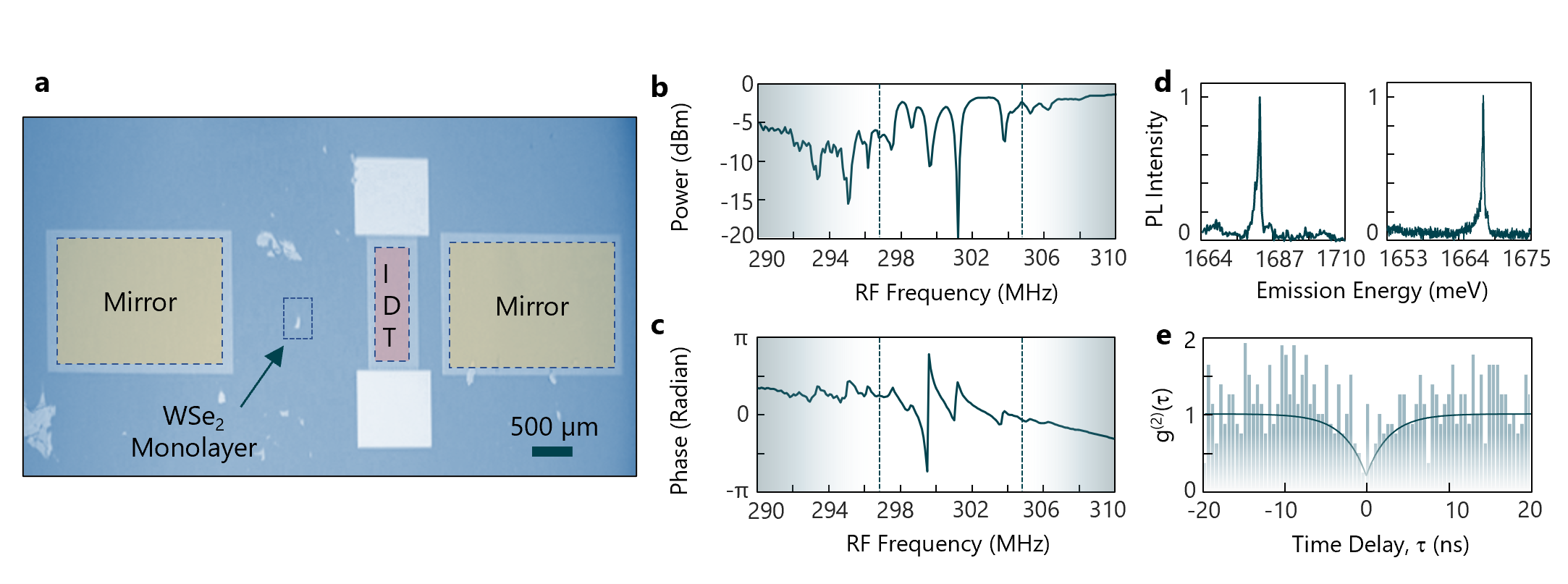}  
          \caption{\textbf{LiNbO$_3$ SAW integration with WSe$_2$ single photon emitters.} \textbf{(a)} Optical image of a 300 MHz SAW cavity with length of 2600 $\mu$m after transfer of a monolayer WSe$_2$. \textbf{(b)} $|S_{11}|$ cavity reflection spectrum showing modes centered at 300 MHz. The gradient areas and vertical dashed lines denote the Bragg mirror band edges. \textbf{(c)} $S_{11}$ phase measurement providing information on the coupling regime of the resonances. Average intrinsic/extrinsic loaded quality factors of 1,900/3,700 were extracted from a simultaneous fit to the magnitude and phase of $S_{11}$. \textbf{(d)} Representative PL spectra of SPEs in the WSe$_2$ monolayer measured at 4.4 K. \textbf{(e)} Second-order auto-correlation function demonstrating photon anti-bunching from an emitter in WSe$_2$ with $g^{\left(2\right)}\left(0\right)$ as low as 0.22.}   
          \label{fig:SAW Integration}
\end{figure*}

In this work, we parametrically modulate the resonance frequency of SPEs in monolayer WSe$_2$ integrated with a LiNbO$_3$ SAW resonator driven by superconducting electronics. We demonstrate cavity phonon-SPE coupling with deformation potential coupling of at least 30 meV/$\%$. The dynamics of the SPE-SAW cavity system is measured through time-resolved, stroboscopic, and steady-state photoluminescence spectroscopy. By sweeping the SAW frequency and the cavity phonon occupation, we demonstrate exquisite control over the local strain. We show that when driven on resonance, the SAW modulates and mixes the emission from exciton fine-structure transitions that have been attributed to anisotropic exchange \cite{he2015single} and intervalley symmetry breaking in the presence of defects \cite{linhart2019localized}, providing a dynamical on-chip control knob for mixing of the WSe$_2$ SPE doublets. These results establish a new experimental platform that combines cavity optomechanics with 2D material quantum optics, paving the way for efficient and high-speed manipulation of 2D quantum emitters for single-photon switching, tuning and stabilization, and entangled photon-pair generation.


\section{Results}

\begin{figure*}[t] \centering
     \includegraphics[scale=1]{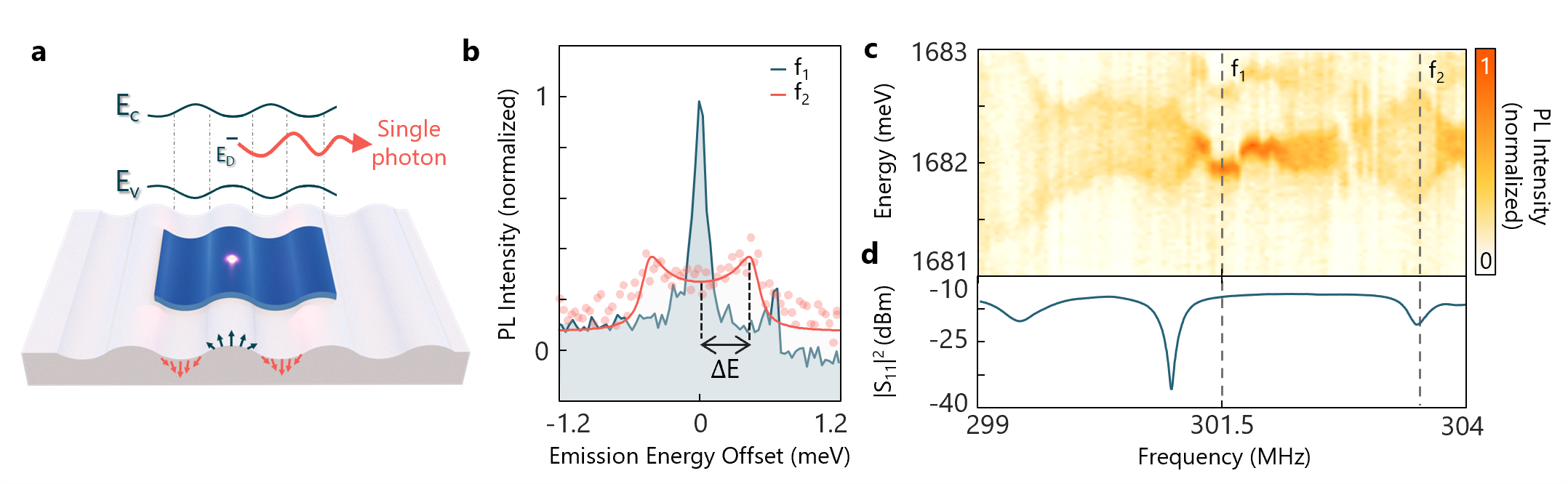}  
          \caption{\textbf{Microwave frequency-dependent energy splitting of a WSe$_2$ single-photon emitter.} \textbf{(a)} Schematic illustration of how the energy bandgap is modulated by the SAW at a given instance in time. The out-of-plane strain vector is indicated by the arrows on the surface of the substrate. \textbf{(b)} Slice plot from \textbf{(c)} showing the single-photon emitter being modulated by the SAW with zero splitting of the emission energy at 301.5 MHz and 0.46 meV splitting at 303.5 MHz \textbf{(c)} The quantum emitter emission modulation as a function of applied frequency to the SAW cavity. The spectral jumps are due to spectral jitter in WSe$_2$ emitters that appear at slow timescales of the measurement under non-resonant optical excitation.  \textbf{(d)} SAW cavity reflection spectrum magnitude corresponding to the modulated emission shown in \textbf{(c)}.}   
          \label{fig:Energy splitting}
\end{figure*}

SAW cavities were fabricated on bulk LiNbO$_3$ using a combination of NbN sputtering and electron beam lithography to define superconducting Bragg reflectors and interdigital transducers (IDTs). The periodicity of the mirrors and the acoustic impedance within the mirrors defines the mirror reflection spectral window, which is centered at 300 MHz for a variety of lengths spanning 900 $\mu$m to 2600 $\mu$m. Electromechanical measurements of the $S_{11}$ parameter for the cavity yield internal quality factors on the order of 12,000 and external quality factors of 7,000, which are limited by the loss due to the mirrors. Using an all dry-transfer technique, monolayer WSe$_2$ was exfoliated and transferred to the SAW cavities \cite{castellanos2014deterministic}. Figure \ref{fig:SAW Integration}(a) shows the optical image of an exemplary SAW resonator and integrated WSe$_2$ flake. $S_{11}$ measurements taken at 4.4 K confirm the presence of the cavity modes, which appear as a series of dips in the cavity reflection spectrum shown in Fig. \ref{fig:SAW Integration}(b). A simultaneous fit to the magnitude and phase (Fig. \ref{fig:SAW Integration}(c)) demonstrates that the cavity is in the undercoupled regime with average intrinsic quality factor of 1900 and extrinsic quality factor of 3700 after the flake transfer, which are similar to loaded quality factors of SAWs integrated with III-V QDs \cite{nysten2020hybrid}.

WSe$_2$ SPEs are first characterized using steady-state photoluminescence (PL) spectroscopy to identify individual emitters, with two representive SPEs shown in Fig.\ref{fig:SAW Integration}(d). Figure \ref{fig:SAW Integration}(e) shows the second-order autocorrelation measurement of an SPE with $g^{\left(2\right)}(0) = 0.22$ demonstrating the anti-bunching behaviour of the emitted light. After identifying the SPE and SAW cavity resonances, the SAW frequency was fixed at the location of each cavity resonance, and the PL measurement was repeated. When driving the SAW on resonance, a standing surface acoustic wave is formed inside of the cavity. Depending on the position of the SPE relative to the node and anti-node of the standing wave, the SPE experiences dynamic compressive and tensile strain from the SAW (Fig. \ref{fig:Energy splitting}(a)). Through deformation potential coupling, the strain modulates the local bandgap of the material, resulting in a temporally varying energy shift of the SPE at the frequency of the SAW mode. Figure \ref{fig:Energy splitting}(b) shows how the emission from an SPE is impacted by the SAW cavity mode. When the cavity mode is populated by driving the IDT on one of the SAW resonances, the SPE zero-phonon line (ZPL) is split into a double-peak structure with a peak-to-peak separation of $2\Delta E = 0.92$ meV. This split-peak structure is a clear signature of SPE-SAW coupling, consistent with previous observations from III-V QDs \cite{Gell2008} and SiC vacancy centers \cite{Whiteley2019} coupled to surface acoustic waves cavities. The PL signals were fit to a Lorentzian function modulated by a sinusoidal interaction in the time domain to extract the modulation energy $\Delta E$ \cite{Manenti2016}. 

Next, the evolution of the PL spectrum as a function of the SAW cavity frequency at a fixed power of 4 dBm was measured (Fig. \ref{fig:Energy splitting}(c)). Maximum modulation of the PL lineshape is evident at the SAW cavity resonant frequencies (Fig. \ref{fig:Energy splitting}(d)) between 299-304 MHz. For the three resonances at 299.425 MHz, 300.975 MHz, and 303.561 MHz, clear splitting is observed, distinctly different from when driving the SAW cavity off resonance. Interestingly, this emitter shows a non-zero splitting between 299-301.5 MHz, which may be attributed to mode mixing between the neighboring cavity resonances due to the SPE being spatially located somewhere between a node and anti-node for this frequency range as similarly reported for QDs \cite{Nysten2020}. Note that among the emitters that were measured, only ones with linewidth narrower than 1 nm (limited by spectrometer resolution) exhibited a measurable splitting. Among these emitters, nearly 75$\%$ exhibited coupling to at least one of the cavity modes, a result which is not surprising given that the positions of the SPEs with respect to node/anti-node of the cavity is randomly distributed. Notably, the SPE in Fig. \ref{fig:Energy splitting}(b) consists of a doublet with a fine-structure splitting on the order of 700 $\mu$eV, which has been previously attributed to anisotropic strain \cite{he2015single} and intervalley excitonic mixing in WSe$_2$ \cite{linhart2019localized,chakraborty2019electrical,parto2021defect}. Interestingly, both peaks of the doublet exhibit the same splitting as a function of frequency, and given the extent of the modulation, the two peaks can overlap and mix at the position of the cavity resonances. This mixing has implications for photon-state engineering and entanglement as discussed below.

\begin{figure*}[t] \centering
     \includegraphics[scale=1.1]{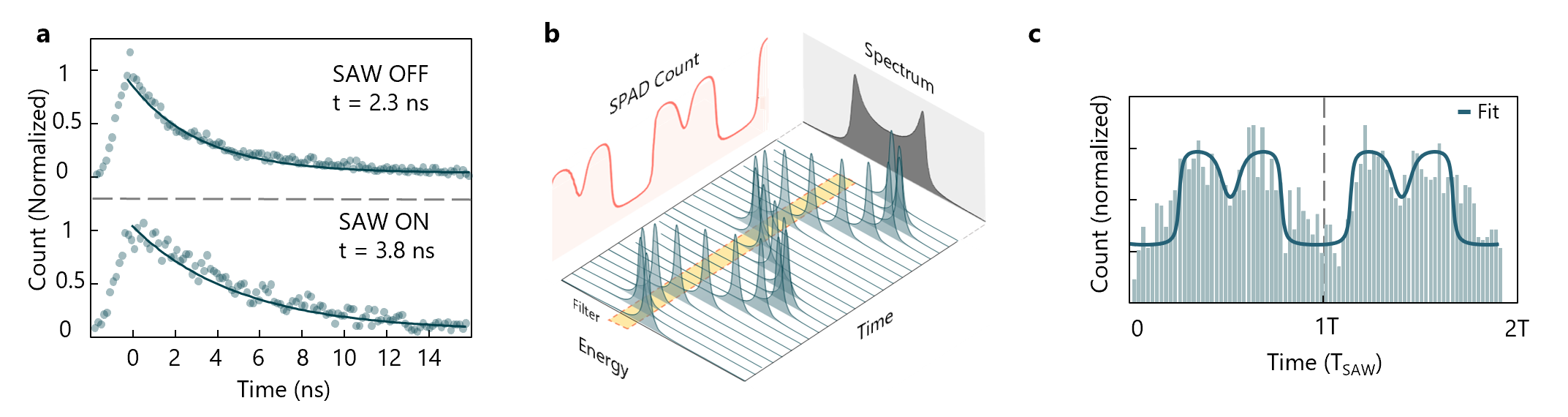}  
          \caption{\textbf{Time-resolved and stroboscopic photoluminescence measurements.} \textbf{(a)} Time-resolved photoluminescence of a WSe$_2$ emitter with the SAW off (top panel) and SAW on (bottom panel). A recombination lifetime > 2 ns is observed in both cases, ruling out any non-radiative, electrodynamic charging, or thermal dissipation mechanisms for the observed modulated signals. \textbf{(b)} Conceptual illustration of the stroboscopic measurement. A monochromator is used to filter out a portion of the modulated signal. The filtered photons are sent to a single photon avalanche diode (SPAD) and their arrival with respect to the SAW drive signal, which modulates the SPE frequency, is recorded. This allows unravelling of the temporal dynamics of the SPE modulation at nanosecond time scales (dark green peaks) which is otherwise inaccessible due to slow time-scale of steady-state PL measurement (dark gray time-averaged spectrum, which is a projection along the time-axis). \textbf{(c)} Results from the stroboscopic measurement (points) with a fit from a Monte Carlo simulation (solid line). Based on the center of the bandpass optical filter, both f$_{RF}$ and a $2f_{RF}$ components are observed as expected.}
          \label{fig:stroboscopic}
\end{figure*}

While it is clear that the SPEs are modulated by the SAWs, to rule out alternative explanations, such as induced non-radiative decay or local heating, we perform time-resolved and stroboscopic PL measurements to map out the SAW dynamics. First, Fig. \ref{fig:stroboscopic}(a) shows the time-resolved PL dynamics of the emitter shown in Fig. 2 with power ($P_{RF} = 6$ dBm) and without power applied to the SAW IDT. In both cases, the SPE recombination lifetime is longer than $\sim 2$ ns, ruling out any new non-radiative recombination or thermal processes that could lead to faster recombination and broadening of the linewidth in the steady-state PL spectrum. We next performed stroboscopic measurements in which the time arrival of the emitted photons with respect to the phase of the applied SAW waveform is measured through single-photon counting and binning. The emission was spectrally filtered using a monochromator to isolate photons near the wings of the modulated steady-state PL spectrum (Fig. \ref{fig:stroboscopic}(b)). Results from this measurement are shown by the histogram in Fig. \ref{fig:stroboscopic} (c), which demonstrates clear modulation of the emission waveform at the fundamental SAW frequency of $f_{RF}$. We observe an additional frequency component at $2f_{RF}$ due to the limited resolution of our monochromator with respect to the total linewidth of our modulated emitter. A fit of the data using a Monte Carlo-like simulation of a modulated emitter overlapped with a non-ideal bandpass filter is shown as the solid line in Fig. \ref{fig:stroboscopic}(c) (see methods).

\begin{figure*}[t] \centering
     \includegraphics[scale=1]{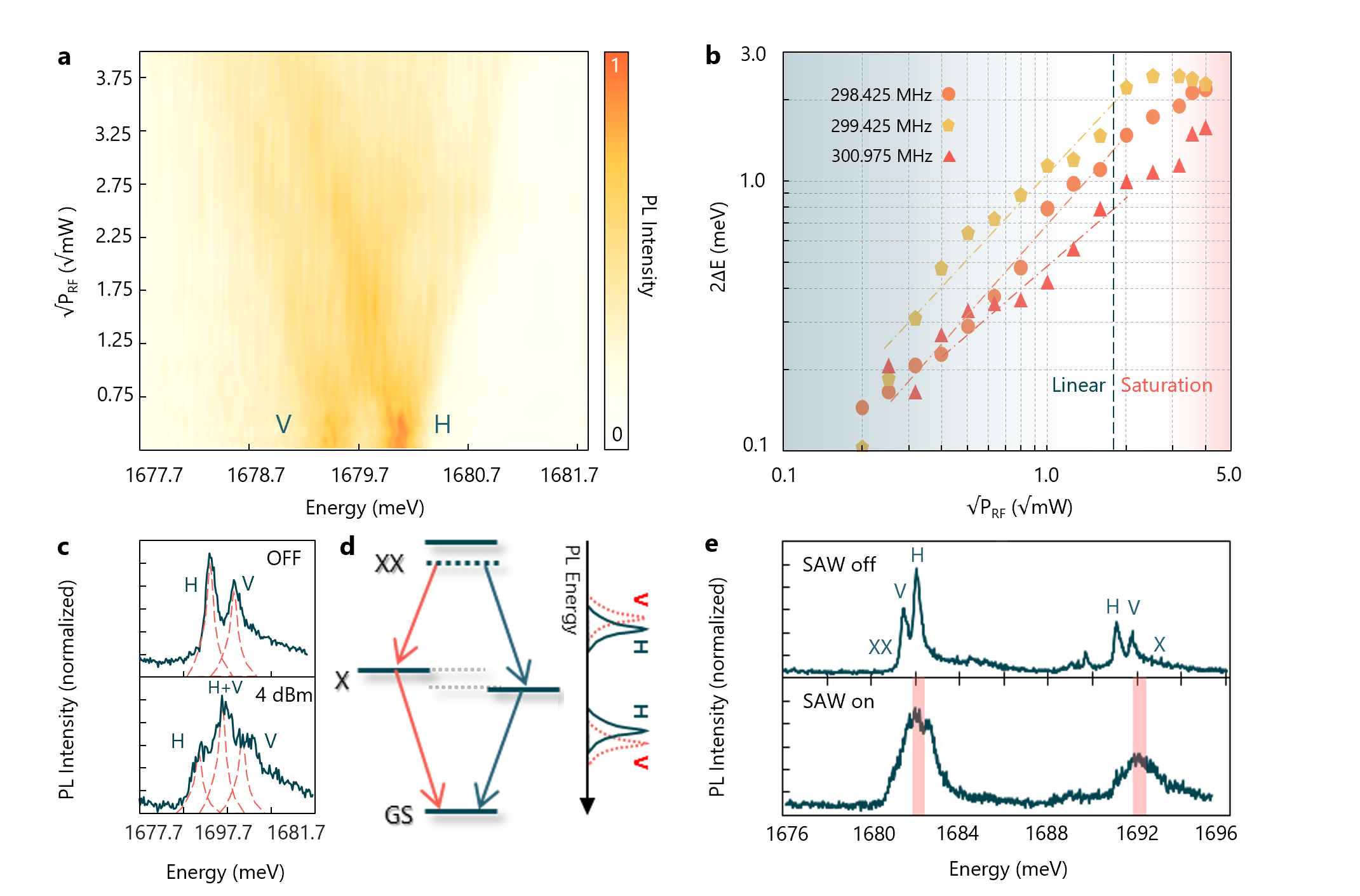}  
          \caption{\textbf{Deformation potential coupling and fine-structure mixing of the exciton-biexciton radiative cascade in monolayer WSe$_2$.} \textbf{(a)} Contour map showing the increase in $\Delta E$ with applied IDT power to a SAW cavity mode at 298.425 MHz. \textbf{(b)} $\Delta E$ as a function of applied power. The average slope from the fits is 0.9865 meV/$\sqrt{mW}$, consistent with deformation potential coupling. \textbf{(c)} Fine-structure splitting of a representative single-photon emitter in WSe$_2$ with horizontally (H) and vertically (V) polarized transitions split into a doublet separated by $0.7$ meV. \textbf{(d)} Mixing of the single-photon emitter fine-structure states in \textbf{(c)} when injecting phonons into the SAW mode at 299.452 MHz. The characteristic double-peak spectrum of SAW-modulated SPEs is observed for both the H and V transitions, which results in a single three-fold Lorentzian in which the central peak arises from mixing of the H and V transitions. \textbf{(d)} Conceptual illustration of biexciton-exciton radiative cascade from WSe$_2$ emitters with non-zero fine-structure splitting. GS, X, and XX denote the ground-state, exciton, and biexciton states. \textbf{(e)} Mixing of the emission from fine-structure states of the exciton-biexciton-like cascade in WSe$_2$ upon SAW modulation. The top panel shows the emission spectrum for no applied SAW power with the characteristic set of doublets corresponding to the exciton-like (X) and biexciton-like (XX) transitions. The bottom panel shows the spectrum from the same emitter with SAW modulation. The doublets merge as demonstrated in \textbf{(c)}, and the highlighted region demonstrates the energies at which the fine-structure splitting is erased by mixing both the H and V photons.}   
          \label{fig:Power dependence}
\end{figure*}

To understand the origin of the SAW-mediated modulation of the SPEs, we next perform experiments measuring $\Delta E$ as a function of the applied power to the IDT. A contour map of the PL spectrum as a function of the square root of the applied power ($P_{RF}$) is shown in Fig.\ref{fig:Power dependence}(a). The splitting $\Delta E$ is extracted at each applied power and fit as previously discussed for each of the resonant cavity modes at 298.425 MHz, 299.425 MHz, and 300.975 MHz. From Fig. \ref{fig:Power dependence}(a), it is clear that $\Delta E$ increases monotonically with the power. On a double-logarithmic scale (Fig.\ref{fig:Power dependence}(b)), we find that $\Delta E \propto \sqrt{P_{RF}}$ with an average slope of 0.9865 meV/$\sqrt{mW}$ for all resonant cavity modes. This result is consistent with deformation potential coupling as the physical mechanism between the SAW and SPE that gives rise to the energy modulation \cite{Pustiowski2015}. At applied power above $\sim$4 dBm, $\Delta E$ saturates, likely due to heating of the NbN IDT and mirrors, which introduces ohmic losses by driving the cavity from the superconducting to normally conducting state. In addition, prior to the saturation regime, the linear fit of the power dependence allows us to rule out any additional contributions, such as Stark-induced electric field coupling, as having a negligible effect on the SPE splitting and modulation, since this would scale as $\Delta E \propto P_{RF}$.

To extract the deformation potential coupling efficiency, a finite element simulation is used to extract the tensile strain component along the propagation axis of the cavity at various applied powers to the IDT. A maximum strain of 0.03$\%$ at 10 dBm applied power is determined for our 300 MHz resonator with a cavity length L = 2600 $\mu$m (see methods). Assuming that the strain in the SAW fully transfers to the WSe$_2$ flake, this results in $\sim$3 meV/0.1$\%$ frequency shift for WSe$_2$ SPEs; however, we note that this is a lower bound for the sensitivity of the emitter, given that the van der Waals interaction between the monolayer and LiNbO$_3$ surface could potentially reduce the full transfer of the strain to the SPE. Previously reported measurements of the static energy shift versus strain applied to SPEs leads to an average value of ~2 meV/0.1$\%$ shift (up to 12 meV/0.1$\%$ maximum) for WSe$_2$ emitters. These values are within the same order of magnitude as our extracted coupling efficiency, which suggests the strain at the LiNbO$_3$ surface is efficiently transferred to the WSe$_2$ monolayer. 

\section{Discussion}

The zero-phonon line of SPEs in WSe$_2$ typically exhibit a doublet with energy splitting of $\sim0.7$ meV, as shown in Fig.\ref{fig:Power dependence}(b). The doublet is thought to arise from asymmetry in the confining potential, which hybridizes the spin and valley states. This hybridization leads to splitting of the SPE transition into orthogonal linearly polarized transitions shown as H and V \cite{linhart2019localized,chakraborty2019electrical,parto2021defect}. Similar fine-structure splitting effects have been observed in other SPE platforms, most notably III-V QDs \cite{bayer2002fine,seguin2005size,schuck2021single}. In some cases, such as the deterministic generation of linearly polarized photons, fine-structure splitting can be advantageous \cite{wang2019towards,gerhardt2019polarization}, while in others, such as generation of entangled-photon pairs via the biexciton-exciton radiative cascade, the fine-structure splitting can be detrimental by reducing the entanglement fidelity \cite{hudson2007coherence,stevenson2006semiconductor}. Despite observations of the radiative biexciton-exciton cascade in monolayer WSe$_2$ \cite{he2016cascaded}, the non-zero fine-structure splitting has prevented any measurements of polarization entanglement with 2D materials.

SAW control of the single-photon emission a unique, on-chip mechanism to manipulate the fine-structure splitting. As shown in Fig. \ref{fig:Power dependence}(c), each fine-structure peak with its associated polarization can be split into two peaks with an energy spacing tuned by the applied SAW IDT power. In a specific range of powers, the two adjacent peaks of the H and V transitions overlap, resulting in a steady-state PL spectrum consisting of three Lorentzians for which the side peaks are vertically and horizontally polarized and the central peak becomes a mixture of the two polarizations (Fig.\ref{fig:Power dependence}). By aligning an optical bandpass filter to the central peak, the collected photons are reminiscent of a atomic-like emitter without fine structure, and the SAW modulation may serve as a mechanism to erase the fine structure altogether, although at a cost of reduced brightness. The SAW-mediated fine-structure mixing may find immediate applications for entangled-photon pair generation from WSe$_2$ SPEs through the radiative biexciton cascade, whereby the emission of a photon from the biexciton-to-exciton transition leads to a second photon emitted from the exciton-to-ground state transition (Fig. \ref{fig:Power dependence}(d)). Indeed, for many SPEs, we observe exciton-biexciton transitions with fine-structure split doublets, as shown in the top panel of Fig. \ref{fig:Power dependence}(d). When turning on the SAW drive power at 299.425 MHz, the deformation potential mixes the transitions in the central regions highlighted in the bottom panel of Fig. \ref{fig:Power dependence}(e). 

The underlying mechanism of the fine-structure mixing still remains an interesting open question worth future investigation. Two potential mechanisms can be envisioned to cause the mixing. First, similar to Fig. \ref{fig:Energy splitting}a, due to oscillating compressive and tensile strain, the bandgap of WSe$_2$ oscillates at the SAW frequency, causing the atomic-level transitions to renormalize and oscillate as well. In this picture, the atomistic description of the two-level system that is responsible for the fine-structure splitting remains relatively the same, and the two fine-structure peaks oscillate in-phase with one another. This mechanism would only appear to have their fine-structure splitting removed on the central peak due to the slow time-scale of the steady-state PL measurements. For the second possible mechanism, the single-particle wavefunctions of the ground and excited states become compressed and elongated across the strain axis, which also causes the fine-structure transition energies to oscillate. Here, however, the fundamental mechanism of the oscillation is not bandgap renormalization, but instead strain-induced modification to the atomic morphology of the defect. In this case, similar to control of the fine-structure splitting via strain in III-V QDs \cite{seidl2006effect}, the compressive and tensile strain can restore the system symmetry associated with the exchange interaction, eliminating the fine structure altogether. In the former case, the central Lorentzian appearing in the PL spectra would become a mixed state of H and V polarized light on slow timescales relative to the inverse of the fine-structure splitting. In the latter case, the central Lorentzian is a coherent entangled state. These two scenarios, on a slow time-scale of steady state PL, lead to the same spectral response, and given that their polarimetric density matrices are identical, they cannot be distinguished from each other with polarization state tomography. It is plausible that both mechanisms, each acting with different strengths, are simultaneously in play. 



While the current iteration of the SPE-SAW resonator is in the initial stage of development, with the resonator operating in a fully classical regime, the next generation of devices require reducing the mode volume and increasing the operation frequency to the gigahertz range to reach the so-called side-band resolved regime. This would allow for coherent quantum phenomena to be observed, such optical sideband pumping to herald the generation of single cavity phonons \cite{imany_2022_quantum}, photon-phonon entangled states, and acoustically driven Rabi oscillations \cite{Whiteley2019}. Although this work focuses on classical control of SPEs with SAWs, we are not limited by any intrinsic property of the 2D emitter-cavity system. WSe$_2$ emitters have been observed with linewidths as low as $\sim 1$ GHz using resonance fluorescence measurements \cite{kumar2016resonant}, and recently near transform-limited hBN emitters \cite{dietrich2020solid} have been reported to have linewidths as low as 200 MHz. In order to bring the system to the side-band resolved regime, the main challenge is to resonantly excite the system to minimize the pure dephasing and to increase the SAW resonance frequencies while maintaining a large \textit{Q} to increase zero-point strain amplitude, bringing the system to the regime of quantum optomechanics \cite{choquer2022quantum}.

\section{conclusion}

In summary, acoustic control of single-photon emitters in monolayer WSe$_2$ integrated with LiNbO$_3$ surface acoustic wave resonators is demonstrated through electro-mechanical and opto-mechanical spectroscopy. The observed single-photon emitter modulation is consistent with deformation potential coupling through strain with sensitivity up to 30 meV/\%. We demonstrate a near-term application of classical control of WSe$_2$ emitters through SAW-mediated single-photon frequency modulation and high-speed fine-structure manipulation, which may open the door for demonstrations of entangled-photon pair generation from 2D materials. The integration of 2D materials with gigahertz-frequency SAW resonators in the future would enable operation in the quantum regime with demonstrations of sideband-resolved excitation and detection, quantum transduction, and photon-phonon entanglement.
\vspace{10pt}

\noindent\textbf{Acknowledgements.} We gratefully acknowledge support from NSF Award No. ECCS-2032272 and the NSF Quantum Foundry through Q-AMASE-i program Award No. DMR-1906325. We thank Hubert Krenner, Matthias Wei\ss, and Emeline Nysten at WWU M\"{u}nster and Kien Le at UCSB for valuable input and discussions.\\

\noindent\textbf{Disclosure.} The authors declare no conflicts of interest. \\

\noindent\textbf{Author Contributions.} S.D.P., K.P., and M.C. contributed equally to this work. G.M conceived the experiments and supervised the project. M.C. designed and fabricated the surface acoustic wave resonators. K.P., S.U., and L.H prepared the samples. S.D.P., M.C., and D.P. assembled the samples. K.P. and S.D.P. performed steady-state PL spectroscopy, time-resolved PL spectroscopy, second-order auto-correlation, stroboscopic PL measurements, and S-parameter measurements. All authors discussed the results and commented on the manuscript at all stages. \\

\section*{Methods}

\subsection*{1. Device Fabrication}

Surface acoustic wave (SAW) resonators were fabricated on bulk 128\textdegree\, YX-cut lithium niobate, which is a piezoelectric substrate with high electromechanical coupling ($K^2 = 5.4\%$). The SAW resonators were fabricated with 20 nm of NbN deposited by DC magnetron reactive sputtering. Distributed Bragg grating mirrors were then patterned using optical lithography and inductively coupled plasma reactive ion etching with CF$_4$ chemistry. Various cavity lengths $L$ were fabricated, where $L = d + 2L_m$. The inner SAW mirror edges are separated by $d$ and the modes penetrate the mirrors by $L_m = w/r_s \approx 130$ $\mu$m for a NbN width $w = 10$ $\mu$m and single-period reflectivity $r_s \approx 0.02$ \cite{nysten2020hybrid, morgan_2007}. One-port SAW resonators were fabricated by placing a NbN interdigital transducer (IDT) within the SAW resonator. Contact pads composed of 10 nm of Ti and 90 nm of Au were deposited using a lift-off process. Finally, monolayer WSe$_2$ flakes were identified using mechanical exfoliation and high-contrast optical imaging. Monolayers were then integrated within the SAW resonator using an all-dry transfer method. After the device fabrication was completed, the samples were attached to an OFHC copper mount that holds both the sample and PCB, and the devices were wire-bonded prior to the experiments.

\subsection*{2. Electro-Mechanical Characterization}

The $S_{11}$ scattering parameter was measured to ascertain the SAW resonator's intrinsic quality factor, $Q_i$, and external quality factor, $Q_e$. The output from a vector network analyzer (VNA) was sent to the single port of the SAW resonator IDT. The reflected signal from the resonator was sent back to the VNA, and the magnitude and phase of $S_{11}$ were measured as a function of RF frequency at 4.4 K. Figure \ref{fig:SAW Integration} shows representative results from this measurement. Dips in the $|S_{11}|$ spectrum within the bandwidth of the SAW resonator mirrors ($\sim 298 - 304$ MHz) are indicative of the different SAW resonator modes. By simultaneously fitting the magnitude and phase at each resonance frequency to Eqn. \ref{Equ:S11}, we extract both $Q_i$ and $Q_e$ for each resonance. 

\begin{equation}
	\label{Equ:S11}
	S_{11}(f)=\frac{(Q_{\mathrm{e},n}-Q_{\mathrm{i},n})/Q_{\mathrm{e},n}+2iQ_{\mathrm{i},n}(f-f_n)/f}{(Q_{\mathrm{e},n}+Q_{\mathrm{i},n})/Q_{\mathrm{e},n}+2iQ_{\mathrm{i},n}(f-f_n)/f}
\end{equation}

\noindent The table below summarizes the results from the fits for each of the cavity resonances measured after transferring the 2D flakes.

\begin{table}[ht]
\centering
\begin{tabular}{|l|l|l|l|l|}
\hline
$f_{SAW}$ [MHz] & 298.425 & 299.425 & 300.975 & 303.561 \\
\hhline{|=|=|=|=|=|}
$\hspace{0.7cm} Q_i$ & \hspace{1.5pt} 1,300 & \hspace{1.5pt} 3,000 & \hspace{1.5pt} 1,600 & \hspace{1.5pt} 1,700 \\
\hline
$\hspace{0.7cm} Q_e$ & \hspace{1.5pt} 5,900 & \hspace{0.15cm} 800 & \hspace{1.5pt} 2,300 & \hspace{1.5pt} 
6,000 \\
\hline
\end{tabular}
\caption{\label{tab:QualityFactors} Intrinsic (Q$_{i}$) and external (Q$_{e}$) quality factors for each of the SAW resonator modes.}
\end{table}


\subsection*{3. Acousto-Optical Microscopy}

A 532 nm continuous-wave laser source was used for the steady-state and stroboscopic measurements. A dichroic mirror at 540 nm is used to separate the optical excitation and collection paths. An additional 600 nm long-pass optical filter is used to further extinguish the excitation laser in the collection path. An infinity-corrected 0.55 NA objective with 13 mm working distance is used for spectroscopy. Samples were placed on a customized radio-frequency (RF) PCB sample carrier for microwave connection and were cooled to 4.4 K inside a Montana S200 cryostation. Optical spectra were acquired using a Princeton instruments HRS-500 with 300/1200/1800 groove/mm gratings and a thermo-electrically cooled Pixis silicon CCD. Second-order autocorrelation measurements with continuous-wave optical excitation were performed by utilizing the spectrometer as a monochromator to filter the emission from individual emitters. The optical signals were then collected in a multimode fiber beamsplitter connected to two single-photon avalanche detectors (Excelitas SCPM-AQRH-13-FC). Swabian time-tagging electronics were used for photon counting. 

Time-resolved photoluminescence measurements were performed using a 660 nm, 80 MHz repetition rate pulsed laser source and single-photon counting. For stroboscopic measurements, the internal oscillator of the RF signal generator for driving the SAW IDT was connected to an external clock generator for synchronization between the SAW and detected photons. The external clock provided a pulsed signal with f$_{RF}$/30 to the start channel of the photon counting module. The emitter photoluminescence detected after the monochromator with the single-photon detector was connected to the stop channel, and a histogram of start-stop times was constructed as the spectrometer grating was scanned across the modulated SPE resonances.\\

\subsection*{4. Time-Domain SAW Simulations}
To understand and predict the temporal dynamics of the stroboscopic measurements, we carried out temporal simulations in MATLAB. In the classical regime, the spectral response of the system at any given time can be denoted as: 

\begin{equation}
    \begin{aligned}
        \Omega(t) = \frac{\Gamma}{1+ \frac{(\omega-\omega_{\circ}-\Delta E sin(2\pi f_{RF}))^2}{\Gamma^2}} 
    \end{aligned}
\end{equation}

where $\Gamma$ and $\omega_{\circ}$ are the radiative decay rate and the frequency of the emitter and $\Delta E$ and $f_{RF}$ are the amplitude of the SAW modulation and frequency of the SAW. The monochromator is modeled as a square pulse in frequency space as $U(\omega_{l})-U(\omega_{h})$ where $\omega_{l}$ and $\omega_{h}$ are the low-pass and high-pass corner frequencies of an ideal bandpass filter. While performing the stroboscopic measurment, the counts on the single photon detector would follow $\Omega(t)(U(\omega_{l})-U(\omega_{h}))$. The fit to the data is calculated using this expression for an emitter with a lifetime of 2 ns, a non-radiatively broadened linewidth of 2 meV, and an ideal bandpass filter with a bandwidth of 3 meV where the filter is set on the high energy wing of the spectral response.

\subsection*{5. Finite Element Simulations of SAW Cavity}

A COMSOL Multiphysics finite-element simulation was used to model the strain amplitude at the location of the WSe$_2$ emitter. A two-dimensional simulation was constructed representing a cross section of the LiNbO$_3$-NbN SAW cavity along the propagation direction and the surface normal. Material parameters for LiNbO$_3$ were extracted from Ref. \cite{tarumi_low_2012}, and the elastic, piezoelectric, and permittivity tensors were then rotated 38\textdegree\, to simulate a 128\textdegree\, YX-cut LiNbO$_3$ substrate. NbN reflectors and IDT electrodes were simulated with electrostatic floating-potential and terminal and ground boundary conditions, respectively. The coupled piezoelectric equations of motion were solved using a frequency-domain simulation with 0 dBm power applied to the IDT electrodes. Perfectly matched layers were applied to the boundaries of the simulation domain to absorb any scattered radiation. The strain amplitude was extracted by taking the maximum value of the tensile strain component oriented along the SAW propagation direction between the IDT and the NbN reflector placed further from the IDT. Over the frequency range of 299-301 MHz, a maximum tensile strain amplitude of 0.0119\,\% at 0 dBm applied IDT power coincided with a SAW cavity resonance at 299.665 MHz, in good agreement with the observed resonance at 299.425 MHz from the $S_{11}$ parameter measurement.

\bibliography{Biblio}
\end{document}